\documentclass[aps,prx,groupedaddress,twocolumn,notitlepage,10pt]{revtex4-1}
\usepackage[utf8]{inputenc}
\usepackage[T1]{fontenc}
\usepackage{graphicx}
\usepackage{subcaption}
\usepackage{amsmath,amssymb,amsfonts}
\usepackage{booktabs}
\usepackage{algorithm}
\usepackage{subcaption}
\usepackage{algpseudocode}

\usepackage{physics}
\usepackage{mathtools}
\usepackage{hyperref}
\usepackage{float}
\restylefloat{algorithm}

\begin{document}
\title{Optimized Quantum Circuit Partitioning Across Multiple Quantum Processors}
% \author{Eneet Kaur, Hassan Shapourian, Jiapeng Zhao, Michael Kilzer, Ramana Kompella, Reza Nejabati}
% \title{Quantum Data Center Infrastructures: A Scalable Architectural Design Perspective}

\author{Eneet Kaur}
\author{Hassan Shapourian}
\author{Jiapeng Zhao}
\author{Michael Kilzer}
\author{Ramana Kompella}
\author{Reza Nejabati}

\affiliation{Cisco Quantum Labs, Santa Monica, CA 90404, USA}

\begin{abstract}
    This paper addresses the challenge of scaling quantum computing by employing distributed quantum algorithms across multiple processors. We propose a novel circuit partitioning method that leverages graph partitioning to optimize both qubit and gate teleportation, minimizing the required Einstein-Podolsky-Rosen (EPR) pairs for executing general quantum circuits. Additionally, we formulate an integer linear program to further reduce entanglement requirements by mapping the logical resources of partitioned circuits to the physical constraints of the quantum network. Finally, we analyze the entanglement cost of implementing the Quantum Fourier Transform (QFT) across multiple QPUs, exploiting the circuit’s structure to minimize total entanglement consumption.
\end{abstract}
\maketitle

Quantum computing has emerged as a promising candidate for solving hitherto complex problems. Indeed, certain quantum algorithms, such as Shor's algorithm \cite{Shor1997} and the HHL algorithm \cite{HHL}, are strongly believed to give a speed-up compared to their classical counterparts \cite{classical_integer_factor}. Application of quantum algorithm range from physical simulations and optimization problems to finance \cite{Lo2016,Peruzzo2014,Shor1997,UrRasool2023,Woerner2019}. In recent years, monolithic quantum computers, i.e. a single quantum processor, has shown promising results \cite{Arute2019,Decross2024}, in sampling problems. However, implementing quantum algorithms that surpass the capabilities of existing classical computers would require coherent control of a large number of qubits.

Extending a monolithic quantum processor to accommodate such a large number of qubits poses significant challenges due to issues such as qubit decoherence, cross-talk, error rates, cooling constraints, and physical layout constraints \cite{manetsch2024tweezerarray6100highly, Stephenson2020}. Distributed quantum computing offers a promising solution to these problems by connecting multiple smaller Quantum Processing Units (QPUs) through a quantum network, as evidenced by the roadmaps of industrial enterprises like IBM \cite{ibm_quantum_roadmap} and IonQ \cite{ionq_quantum_networks}. Distributed quantum computing also offers the advantage of modularity—once the limitations of a particular quantum processing platform chip is reached, an extra unit can be added to the existing infrastructure.

Quantum circuits are composed of a sequence of single- and two-qubit gates, measurements, and classical communication \cite{Kitaev1997}. The implementation of the two-qubit gates can vary accross platforms \cite{Veldhorst2015, Sackett2000,Evered2023,ion_trap_two_qubit}. In distributed quantum computing, two-qubit gates become even more challenging. These gates can be divided into local or non-local gates, with local gates implemented within a single QPU and non-local gates implemented between qubits in different QPUs. Establishing non-local gates between QPUs necessitates a quantum communication link between the quantum processors to establish EPR pairs.  

%Superconducting qubit platforms implement two-qubit gates, like the CNOT gate, via cross-resonance (CR) gates, where one qubit is driven at the frequency of another to create an interaction \cite{Veldhorst2015}. Trapped ion platforms use laser beams to manipulate ions, with the Mølmer–Sørensen gate being common, using bichromatic lasers to entangle ions \cite{Sackett2000}. Neutral atom platforms employ optical tweezers and implement two-qubit gates using Rydberg interactions, where atoms excited to high-energy Rydberg states interact strongly, forming controlled phase gates that can be transformed into CNOT gates \cite{Evered2023,ion_trap_two_qubit}. 

%such as photonic interconnects between ion traps \cite{Monroe2013} and shuttling of trapped ions \cite{shuttling_ion}. While numerous candidates exist for implementing non-local gates, they are currently noisier and take more time to implement than local two-qubit gates.

As specified above, a critical resource for executing distributed quantum circuits is the availability of high-fidelity EPR pairs across QPUs. Generating these EPR pairs tends to be both slow and prone to noise, which can degrade the performance of quantum algorithms \cite{Knaut2024,Stephenson2020,Liu2024}. In this work, we seek to minimize the number of EPR pairs or non-local gates required to run a quantum algorithm on multiple processors.

We assume that the QPUs will be equipped to interface with a quantum network, potentially using quantum frequency converters to enable photon-based EPR pairs between QPUs. To this end, some pairs of QPUs could have higher entanglement generation rates and support higher fidelities \cite{DQC_cisco}. We aim to distribute the algorithm in a way that minimizes the entanglement demand on links with higher noise (lower fidelity) and limited qubit transfer bandwidth (lower capacity).  A major distinction between this work and previous studies is that we allow for changes in qubit allocation during the execution of the circuit. The problem statement considered in this work can be framed as follows:

Given quantum processing units equipped with a quantum network and given an arbitrary quantum circuit decomposed into single-qubit gates and two-qubit gates, what should be the qubit allocation during the run of the circuit across QPUs to (a) minimize the overall number of required EPR pairs and (b) minimize the EPR pairs required over low fidelity/capacity links?

Even finding the optimal solution for minimizing the overall EPR pairs required by an algorithm is NP-hard \cite{Andreev2006}. Therefore, to address the question posed above, we introduce the following heuristic:

\begin{itemize} \item We propose an algorithm based on graph partitioning techniques for optimizing the number of EPR pairs required to run a quantum circuit on an arbitrary number of QPUs and qubits. This algorithm optimizes both \textit{qubit} teleportation and \textit{gate} teleportation to realize a two-qubit gate. The output of this algorithm specifies the number of EPR pairs required between any pair of QPUs to execute the transpiled quantum circuit. \item We construct an integer linear program (ILP) to minimize the EPR requirement over the low-fidelity links. The solution to the ILP maps the resources required for the partitioned circuit to the physical resources in the network. \end{itemize}

Next, we analyze the entanglement resources required to implement Quantum Fourier Transform (QFT) on multiple processors. We utilize the structure of the QFT circuit to minimize the total entanglement needed to implement QFT jointly on multiple QPUs.  

\section{Previous works}
Two of the earliest heuristics developed for circuit partitioning were introduced in \cite{ZomorodiMoghadam2017} and \cite{AndrsMartnez2019}. The work in \cite{ZomorodiMoghadam2017} focuses on minimizing the number of qubit teleportations required for a distributed quantum circuit across two partitions. A method that maps quantum circuits to hypergraphs was introduced in \cite{AndrsMartnez2019}, involving two key phases: a preprocessing phase that groups equivalent gates, and a second phase where hypergraph partitioning is performed using KaHyPar\cite{Kahypar}. This approach primarily involves grouping controlled gates with the same control qubit and employing the CAT entangler and disentangler method for implementing non-local gates \cite{cat_entangler}. Genetic algorithms were used in \cite{genetic_circ} to minimize qubit teleportation costs in circuit partitioning. The KL heuristic was utilized in \cite{Daei2020} to perform a $k$-way balanced partition of quantum circuits, where $k \geq 2$. The work in \cite{Nikahd2021} used ILP to incorporate both gate and qubit teleportation. A method for efficiently utilizing entanglement in distributed quantum computing was introduced in \cite{Wu2023}, and this method was later extended to homogeneous networks in \cite{Martinez2023}.

A two-step heuristic was proposed in \cite{Sundarama2021}, where qubits are first partitioned across $k$ near-balanced processors, followed by the use of cat entanglement for non-local gates. This work also introduced a heuristic for minimizing the set of migrations required to implement the non-local gates, and was later extended to include teleportation in \cite{sundaram2022distributionquantumcircuitsgeneral}. The authors in \cite{davarzani2020dynamicprogrammingapproachdistributing} proposed using bipartite graphs and dynamic programming to minimize qubit teleportations. A method to map quantum circuits to modular physical machines, one time slice at a time, was presented in \cite{Baker_2020}, optimizing qubit assignments for each time slice and accounting for the cost of moving qubits from the previous time slice, using a tunable lookahead scheme to reduce future movement costs. This work was then extended in \cite{burt2024} and includes gates packing \cite{Wu2023}, wherein the solution space is explored via genetic algorithms. The Hungarian algorithm was employed in \cite{Escofet2023} to propose a heuristic for the qubit assignment problem. Deep reinforcement learning was utilized for circuit partitioning on multi-core architectures in \cite{Pastor_2024}.

In \cite{yimsiriwattana2004generalizedghzstatesdistributed}, the authors considered the distribution of Quantum Fourier Transform (QFT) across two QPUs, calculating the quantum communication resources for controlled phase gates to be $O(n^2)$. In this work, teleportation was used to implement the swap gates. \cite{Neumann2020} also studied the implementation of QFT between two processors, reducing the number of EPR pairs required for CPhase gates to $n$. However, the swap gates required at the end of the QFT were accounted for through careful qubit allocation bookkeeping.

In recent years, multiple heuristics have been developed to minimize the number of EPR pairs required to execute circuits across multiple quantum cores, often relying on methods such as qubit teleportation, gate teleportation, and packing multiple two-qubit gates to reduce entanglement costs. These approaches typically fix qubit assignments at the start of the circuit, limiting flexibility, and often require two-way teleportation protocols, where qubits must return to their original QPU after operations. In contrast, our work introduces a novel framework that allows for both qubit and gate teleportation with dynamic, one-way teleportation, enabling qubits to be reassigned on the fly without necessitating their return to the originating QPU. This flexibility not only minimizes the entanglement overhead but also enables efficient allocation of non-local gates.

\section{Notation and basics}
A quantum circuit can be decomposed into a sequence of one-qubit and two-qubit gates \cite{Kitaev1997}. One-qubit gates operate on individual qubits and include operations such as rotations and phase shifts. Two-qubit gates, such as the CNOT gates, involve entangling operations between pairs of qubits. In this work, we assume that we are given a decomposition of the quantum circuit.

We use the term \textit{QPU computation capacity} to refer to the number of qubits that a QPU can allocate for computation. Additionally, we assume each QPU is equipped with finite communication qubits. These communication qubits are used to establish entanglement between computation qubits of QPUs. 
Once entanglement is established, it can be used to implement two-qubit gates. Notably, any two-qubit gate can be decomposed into a CNOT gate and single-qubit rotations. Below, we explain how to implement a CNOT gate using a shared EPR pair.

\begin{figure*}
    \centering
    \includegraphics[width=\linewidth]{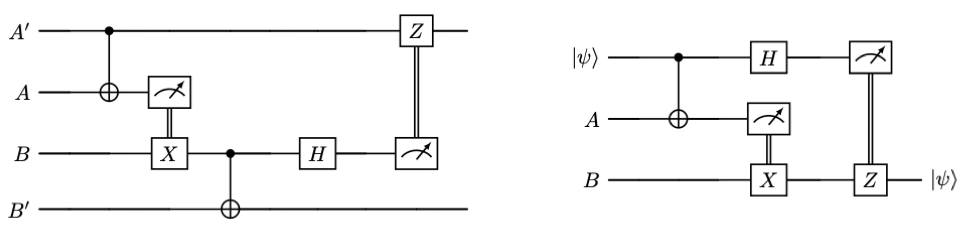}
    \caption{Quantum circuits implementing gate teleportation (left) and state teleportation (right). Qubits $A$ and $B$ share an EPR pair }
    \label{fig:teleportation}
\end{figure*}

 \textit{Quantum state teleportation} \cite{Teleportation} is a technique where the state of a qubit can be transferred to another qubit using quantum entanglement. The qubit in the state $\ket{\psi}$ to be teleported interacts with an EPR pair. Then, one half of the EPR pair and the qubit are measured in the $X$ and $Z$ bases. The state of the other half of the pair becomes a Pauli equivalent of the state $\ket{\psi}$. Once the state of the qubit is teleported to the communication qubit, it is further teleported to the computation qubit. After this process, the qubits within the QPU can locally apply any two-qubit gates. See Fig.~\ref{fig:teleportation} for details. 

\textit{Gate teleportation} \cite{Gottesman1999,PhysRevA.62.052317} is a technique in which a remote CNOT gate is applied between two qubits $A'$ and $B'$ located in separate QPUs without teleporting the state of the qubit to either of the QPUs. The EPR pair is used to implement a non-local CNOT gate. The circuit diagram that implements this method is shown in Fig.~\ref{fig:teleportation}. 

If a circuit has a larger number of upcoming gates where one qubit is consistently involved with different qubits located on another QPU, qubit teleportation (state teleportation) is preferable to gate teleportation. By teleporting the frequently involved qubit to the QPU hosting the other qubits, subsequent two-qubit gates, such as CNOTs, can be executed locally. This reduces the need for repeated inter-QPU communication, enhancing efficiency and resource utilization.

% By leveraging these techniques, we can facilitate the implementation of two-qubit gates across different QPUs, thereby enabling distributed quantum computing. While both qubit teleportation and gate teleportation can be used to implement the CNOT gate, which particular technique to use depends on the structure of the decomposed circuit.

\textit{Graph partitioning} is a fundamental problem in computer science and applied mathematics, where the objective is to divide the vertices of a graph into disjoint subsets, also known as partitions, while minimizing the number of edges or total weights of the edges between different partitions. Graph partitioning problem is a NP hard problem. To solve this problem, there exists a number of heuristics such as KL algorithm \cite{KL_algorithm}, Metis \cite{Karypis1998}, and genetic algorithms \cite{GA_graph_part} among others. The Kernighan-Lin (KL) algorithm is a heuristic method that iteratively refines an initial partition by swapping pairs of vertices between subsets to reduce the edge cut. This method can be highly influenced by the choice of the intitial partition. Another approach METIS is often employed useful for large graphs. 

\textit{Circuit partitioning} can be defined as dividing a circuit for execution across multiple QPUs, where qubits are distributed among different QPUs, and their allocation can dynamically change during the execution of the circuit.  Any circuit partitioning heuristic should address two key aspects: how qubits are allocated to multiple QPUs throughout the circuit’s execution, and which gates are to be executed non-locally.

\begin{figure}[h!]
    \centering
    \begin{subfigure}[b]{0.5\textwidth}
        \centering
        \includegraphics[width=\textwidth]{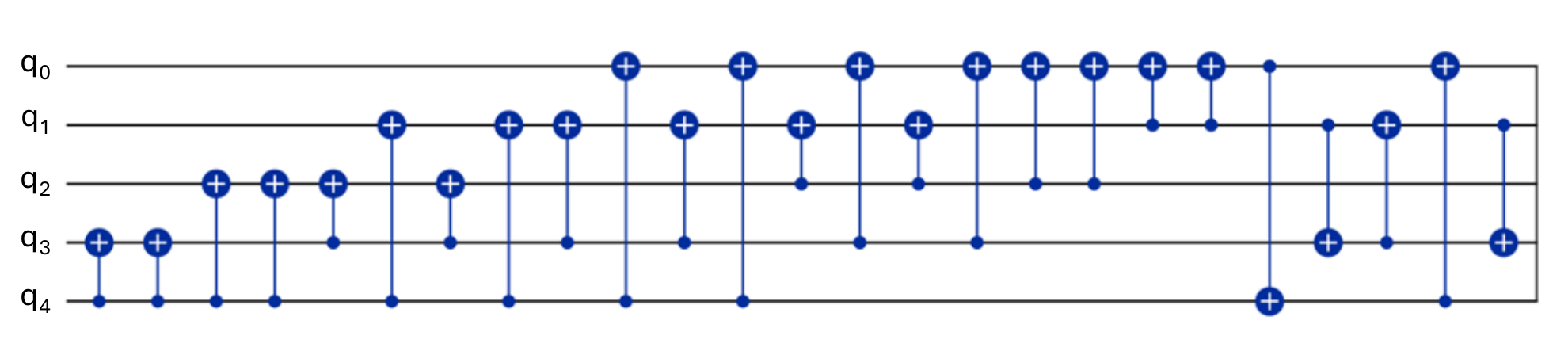} % Replace with your image file
        \caption{Sample circuit}
        \label{fig:sample_circuit}
    \end{subfigure}
    \hspace{1cm}
    \begin{subfigure}[b]{0.3\textwidth}
        \centering
        \includegraphics[width=\textwidth]{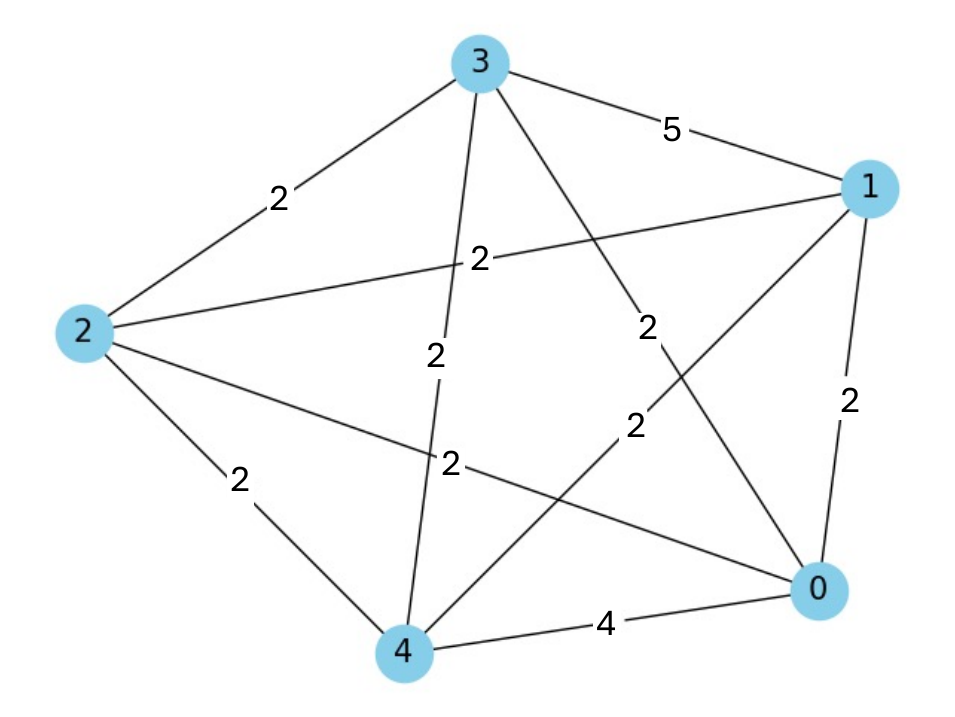} % Replace with your image file
        \caption{Mapping circuit to a graph}
        \label{fig:graph}
    \end{subfigure}
    \caption{Mapping of a quantum circuit to a graph}
    \label{fig:circuit_to_graph}
\end{figure}

The circuit partitioning problem can be simplified and mapped exactly onto a graph partitioning problem under the condition that only gate teleportation is used for implementing non-local CNOT gate. In this case, consider the following mapping: to each qubit $q_i$, where $i \in [1,n]$ and $n$ is the number of qubits in the circuit, denote a node $i$ in graph \( G \). For each CNOT gate between qubits \( q_i \) and \( q_j \), construct an edge between $i$ and $j$ and assign a weight equal to the number of CNOT gates between \( q_i \) and \( q_j \). Then, one can employ any graph partitioning techniques that minimize the weight of the edges between the partitions. For the pseudo code, see Algorithm~\ref{algorithm:gate}. See an example in Fig.~\ref{fig:circuit_to_graph}. While this technique works well for certain circuits, it does not allow for qubit teleportation to implement non-local gates. This can lead to an overestimation of the required EPR pairs.

The Kernighan-Lin (KL) algorithm \cite{KL_algorithm} is a heuristic algorithm used for partitioning graphs (see pseudo code Algorithm~\ref{algorithm:kl}). It aims to minimize the total edge weight between partitions. The original KL algorithm enforces balanced partitions, where the number of nodes in each partition is equal. However, the modified version relaxes this constraint, allowing for partitions with nodes \(n_1\) and \(n_2\), such that \(n_1 + n_2 = n\), where \(n\) represents the total number of nodes in the graph. The algorithm operates iteratively by swapping pairs of nodes between partitions to reduce the total edge weight.

% \subsection*{Steps of the KL Algorithm}

% \begin{enumerate}
%     \item \textbf{Initial Partitioning:}
%     \begin{itemize}
%         \item The graph is initially divided into two equal-sized partitions.
%     \end{itemize}
    
%     \item \textbf{Gain Calculation:}
%     \begin{itemize}
%         \item For each node in both partitions, the algorithm calculates the gain, which represents the reduction in edge cut size if the node were to be moved to the other partition.
%     \end{itemize}
    
%     \item \textbf{Node Swapping:}
%     \begin{itemize}
%         \item Pairs of nodes from each partition are selected and swapped to maximize the gain. The swaps continue until no further improvement can be made.
%     \end{itemize}
    
%     \item \textbf{Convergence:}
%     \begin{itemize}
%         \item The process is repeated iteratively until a local minimum in the edge cut size is reached, meaning no further swaps can reduce the cut size.
%     \end{itemize}
% \end{enumerate}
\begin{algorithm}[H]
\caption{Circuit Partitioning Using Gate Teleportation}
\begin{algorithmic}[1]
\State \textbf{Input:} Quantum circuit with qubits \( q_1, q_2, \ldots, q_n \) and CNOT gates
\State \textbf{Output:} Partitioned graph \( G \) minimizing inter-partition edge weights

\State Initialize an empty graph \( G = (V, E) \)
\For{each qubit \( q_i \) in the circuit}
    \State Add node \( v_i \) to \( V \)
\EndFor

\For{each CNOT gate between qubits \( q_i \) and \( q_j \)}
    \If{edge \( (v_i, v_j) \) already exists in \( E \)}
        \State Increment the weight of edge \( (v_i, v_j) \) by 1
    \Else
        \State Add edge \( (v_i, v_j) \) to \( E \) with weight 1
    \EndIf
\EndFor

\State Employ a graph partitioning technique to partition graph \( G \) into subgraphs \( G_1, G_2, \ldots, G_k \) such that the sum of the weights of the edges between the partitions is minimized
\State \textbf{Return} partitioned graph \( G \)

\end{algorithmic}
\label{algorithm:gate}
\end{algorithm}

\begin{algorithm}[H]
\caption{KL Algorithm for Graph Partitioning}
\begin{algorithmic}[1]
\State \textbf{Input:} Graph \( G = (V, E) \) with nodes \( v_1, v_2, \ldots, v_n \)
\State \textbf{Output:} Two partitions \( P_1 \) and \( P_2 \) minimizing total edge weight

\State \textbf{Step 1: Initial Partitioning}
\State Divide graph \( G \) into two partitions \( P_1 \) and \( P_2 \) with nodes $n_1$ and $n_2$, with $n_1+n_2 =n$. 

\Repeat
    \State \textbf{Step 2: Gain Calculation}
    \For{each node \( v_i \) in partition \( P_1 \)}
        \State Calculate the gain \( g_i \) for moving \( v_i \) to \( P_2 \)
    \EndFor
    \For{each node \( v_j \) in partition \( P_2 \)}
        \State Calculate the gain \( g_j \) for moving \( v_j \) to \( P_1 \)
    \EndFor

    \State \textbf{Step 3: Node Swapping}
    \While{there is a pair of nodes that can be swapped to improve gain}
        \State Select nodes \( v_i \in P_1 \) and \( v_j \in P_2 \) to swap
        \State Swap \( v_i \) and \( v_j \)
    \EndWhile
\Until{no further improvement in edge cut size}

\State \textbf{Return} final partitions \( P_1 \) and \( P_2 \)
\end{algorithmic}
\label{algorithm:kl}
\end{algorithm}

\section{Algorithm -- window based circuit partitioning}
In this section, we introduce a window-based circuit partitioning (WBCP) algorithm that takes as input an arbitrary quantum circuit, the number of QPUs, and the computational capacity of each QPU. The algorithm outputs the qubit allocation during circuit execution, optimizing to minimize the number of EPR pairs consumed between QPUs. As a result, each two-qubit gate in the circuit is assigned either a local or non-local label. The use of qubit teleportation for implementing non-local gates implies that the qubit allocation is \textit{dynamic} and can change during the execution of the circuit.

\subsection{Algorithm Description}

\begin{figure}
    \centering
    \includegraphics[width=\linewidth]{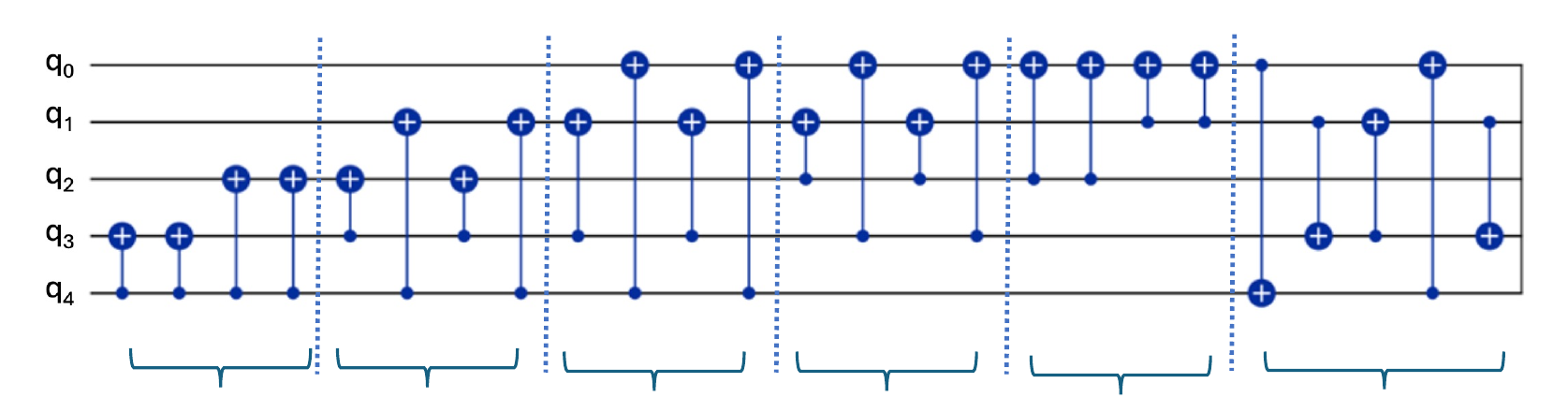}
    \caption{In this figure, we illustrate an application of the WBCP on a five-qubit circuit. The circuit is divided into sub-circuits of length $l$. Within each sub-circuit, remote gates are implemented using gate teleportation. Between sub-circuits, qubits can be rearranged through state teleportation.}
    \label{fig:algorithm_example}
\end{figure}

The crux of the algorithm is to divide the quantum circuit $C$ into sub-circuits $C_i$, within which only gate teleportation is allowed. At the beginning of each sub-circuit, the algorithm looks ahead and determines whether the teleportation of qubits would be useful in minimizing the EPR requirements of that sub-circuit. For an example, see Fig.~\ref{fig:algorithm_example}. In the following text, we define the entanglement cost as a function of the partition $P_k$ and the sub-circuit $C_i$, i.e., \(EC_{P_k}^{(C_i)}\), which is the total edge weights across the partition \(P_k = \{P_k^j\}_{j=1}^m\) of the graph \(G\) corresponding to \(C_i\) and $m$ is the number of disjoint subsets in the partition $P_k$. The mapping of the sub-circuit \(C_i\) to \(G\) is in accordance with the mapping in Algorithm~\ref{algorithm:gate}. The pseudocode of the window-based partitioning algorithm is given in Algorithm~\ref{alg:window}.

\begin{enumerate}
    \item \textbf{Inputs:} The algorithm takes as inputs a quantum circuit \( C \) and a window length \( l \).
    
    \item \textbf{Divide the Circuit:} The circuit \( C \) is divided into smaller sub-circuits \( \{C_i\}_{i=1}^{m} \), where \( m = \lceil n/l\rceil \) and \( n \) is the total number of two-qubit gates in the circuit.
    
    \item \textbf{Initialize the Entanglement Cost:} The process begins by initializing the entanglement cost (EC) to zero. 

Each subcircuit \( C_i \) is processed sequentially. For the first subcircuit (\( i = 1 \)), the algorithm executes Algorithm~\ref{algorithm:gate}, which returns the partition \(\{P_1^j\}_{j=1}^m\).

For every subsequent subcircuit \( C_i \) (\( i > 1 \)), the process involves several steps.  A graph \( G_i \) is constructed, with nodes representing the qubits involved in \( C_i \). Each pair of nodes (qubits) in this graph is assessed: if both nodes belong to the same subset of the previous partition \( P_{i-1} \), the weight is set to twice the number of CNOT gates connecting them; if they belong to different subsets, the weight equals the number of CNOT gates.

A partitioning algorithm is applied to \( G_i \), resulting in a new partition \( P_{new} \). The entanglement costs for the new and old configurations, \( EC_{P_{new}}^{C_i} \) and \( EC_{P_{i-1}}^{C_i} \) respectively, are calculated. The new cost is adjusted by adding the number of qubits moved, calculated with respect to the partitions \( P_{i-1} \) and \( P_{new} \).

The algorithm then compares these costs: if \( EC_{P_{new}}^{C_i} \) is less than or equal to \( EC_{P_{i-1}}^{C_i} \), the new partition \( P_i = P_{new} \) is adopted; otherwise, the partition remains \( P_i = P_{i-1} \). Finally, the smaller of the two entanglement costs is added to the total entanglement cost \( EC \).

    \item \textbf{Output:} The algorithm outputs the total entanglement cost \( EC \), and qubit allocation for each window.
\end{enumerate}

The algorithm is optimized for minimizing the EPR requirements when executing the circuit \(C\), allowing for both qubit and gate teleportation. It is important to note that the outcome of the algorithm is highly dependent on the window length \(l\). To obtain the optimal value, we sweep through various window lengths. When the window length equals the circuit length, the window-based graph partitioning algorithm becomes equivalent to Algorithm~\ref{algorithm:gate}.

Additionally, our algorithm prioritizes keeping qubits in the same partitioning set, as evidenced by the increased edge weights associated with CNOT gates between qubits that belong to the same sets from previous partitions. The increase in weight is designed to encourage qubits to remain in the same partition, while not being drastic enough to completely inhibit teleportation between the windows, thus taking into account the history of the sub-circuits.

\subsection{Graph partitioning techniques}
In both Algorithm~\ref{algorithm:gate}  and \ref{alg:window}, we use graph partitioning algorithms. While any graph partitioning technique can be invoked, we use both the Metis and KL algorithms. The Metis algorithm, while offering an improvement in runtime, has limitations when controlling the computational capacity of the QPUs. At times, the internal structure of the created graph can lead to an imbalanced division. On the other hand, the KL heuristic, although it requires more runtime, provides better control in constraining the computational capacities of the QPUs.

Another benefit of the KL algorithm is that the initial partition in Algorithm~\ref{algorithm:kl} for the sub-circuit \(C_i\) can be set to the optimal partition determined for the previous sub-circuit \(C_{i-1}\). This approach leverages historical partitioning data to minimize the need for teleportation between sub-circuits.

% \begin{algorithm}
% \caption{Graph Partitioning for Quantum Circuits}
% \begin{algorithmic}[1]
% \State \textbf{Input:} Circuit $C$, Window Length $l$
% \State Divide $C$ into subcircuits $\{C_i\}_{i=1}^{m}$, $ m = \lceil n/l\rceil $, $n$ is the number of two-qubit gates.
% \State Initialize $EC = 0$

% \For {each subcircuit $C_i$}
%     \If {$i == 0$}
%         \State To each qubit in $C$ assign a node in $G_0$.
%         \State Set edge weights as number of CNOTs between nodes
%         \State Obtain partition $P_0$ using graph partitioning.
%         \State $EC \mathrel{+}= $ teleportation cost of $P_0$ for $C_0$
%     \Else
%         \State Define graph $G_k$ with nodes from $C_i$
%         \For {each pair $(i,j)$ in $G_k$}
%             \State Set weight $= 2 \times $ \# CNOTs if in same partition, else \# CNOTs
%         \EndFor
%         \State Obtain partition $P_{new}^i$ using partitioning algorithm
%         \State Calculate $EC_{new}^i$ and $EC_{old}^i$ for $C_i$
%         \State $P_i = P_{new}^i$ if $EC_{new}^i < EC_{old}^i$ else $P_i = P_{i-1}$
%         \State $EC \mathrel{+}= \min(EC_{old}^i, EC_{new}^i)$
%     \EndIf
% \EndFor
% \State \textbf{Output:} Total entanglement cost $EC$
% \end{algorithmic}
% \label{algorithm:window_based}
% \end{algorithm}

\begin{algorithm}[H]
\caption{Algorithm for Minimizing EPR Requirements}
\begin{algorithmic}[1]
    \State \textbf{Inputs:} A quantum circuit \(C\) and a window length \(l\).
    
    \State Divide the circuit \(C\) into smaller sub-circuits \( \{C_i\}_{i=1}^{m} \), where \( m = \lceil n/l\rceil \) and \( n \) is the total number of two-qubit gates in the circuit.
    
    \State \textbf{Initialize the Entanglement Cost:} Set the initial entanglement cost \(EC\) to 0.
    
    \For {each subcircuit \( C_i \)}
        \If {\( i = 1 \)} 
            \State Run Algorithm~\ref{algorithm:gate} and obtain the partition \(\{P_1^j\}_{j=1}^m\).
        \Else
            \State Create a graph \( G_i \) where the nodes represent the qubits involved in \( C_i \).
            \For {each pair of nodes (qubits) in \( G_i \)}
                \If {the nodes are in the same subset of the partition \(P_{i-1}\)}
                    \State Set the weight to 2 times the number of CNOT gates between them.
                \Else
                    \State Set the weight to the number of CNOT gates.
                \EndIf
            \EndFor
            \State Apply a partitioning algorithm to partition the graph \( G_i \) into a new partition \( P_{new} \).
            \State Calculate the new entanglement cost \( EC_{P_{new}}^{C_i} \) and the old entanglement cost \( EC_{P_{i-1}}^{C_i} \).
            \State Update \( EC_{P_{new}}^{C_i}  = EC_{P_{new}}^{C_i} + \text{\#qubits moved} \), where qubits moved are calculated with respect to the partitions \(P_{i-1}\) and \(P_{new}\).
            \If {\( EC_{P_{new}}^{C_i} \leq EC_{P_{i-1}}^{C_i} \)}
                \State Set \( P_i = P_{new} \).
            \Else
                \State Set \( P_i = P_{i-1} \).
            \EndIf
            \State Add the smaller of \( EC_{P_{new}}^{C_i} \) and \( EC_{P_{i-1}}^{C_i} \) to the total entanglement cost \( EC \).
        \EndIf
    \EndFor
    
    \State \textbf{Output:} 
    \Statex \quad The total entanglement cost \( EC \), and the qubit allocation for each window.
\end{algorithmic}
\label{alg:window}
\end{algorithm}

% To enhance the capabilities of the KL algorithm for our specific requirements, we have extended the KL heuristic in the following ways:

% First, we have implemented a recursive KL algorithm that generalizes the partitioning process for any number of arbitrary partitions, not just two. This is crucial for distributing quantum circuits on more than two QPUs.

% Second, the imbalanced KL algorithm allows users to specify the number of computational nodes in each Quantum Processing Unit (QPU). This feature provides greater control over the computational capacities of qubits, ensuring that the partitioning process considers the available resources and distributes the workload accordingly.

% By using these extensions, we achieve a more balanced and controlled partitioning of the quantum circuits, which is essential for optimizing the performance of distributed quantum computing systems.

% Note here that unlike previous techniques, we do not require the teleported qubit to be back in the original QPU. Infact, at the end of the window, the qubit if moved from the original QPU stays in the new QPU. This can be especially useful when considering the circuit partitioning in multiple QPUs. 

% The use of the metis algorithm also allows for arbitrary number of partitions.

We have also enhanced the KL algorithm by implementing a recursive version that generalizes the partitioning process to accommodate any number of partitions, rather than being restricted to just two. This enhancement ensures that Algorithm~\ref{alg:window} can distribute quantum circuits across multiple QPUs, similar to the functionality already provided by the Metis algorithm.

Furthermore, we have modified the KL algorithm to support imbalanced partitions, allowing users to specify the number of computational nodes within each Quantum Processing Unit (QPU). This modification provides greater control over the allocation of qubits, ensuring that the partitioning process is tailored to the available computational resources.

Unlike previous methods, our approach does not require teleported qubits to return to their original QPU. Instead, once a qubit is transferred to a new QPU during the partitioning process, it remains there at the end of the window. This strategy is particularly beneficial when partitioning circuits across multiple QPUs.

%Moreover, the use of the Metis algorithm enables partitioning into an arbitrary number of partitions, further enhancing the flexibility and effectiveness of the partitioning process.
\subsection{Numerical results}
\begin{figure}
    
        \includegraphics[width=\linewidth]{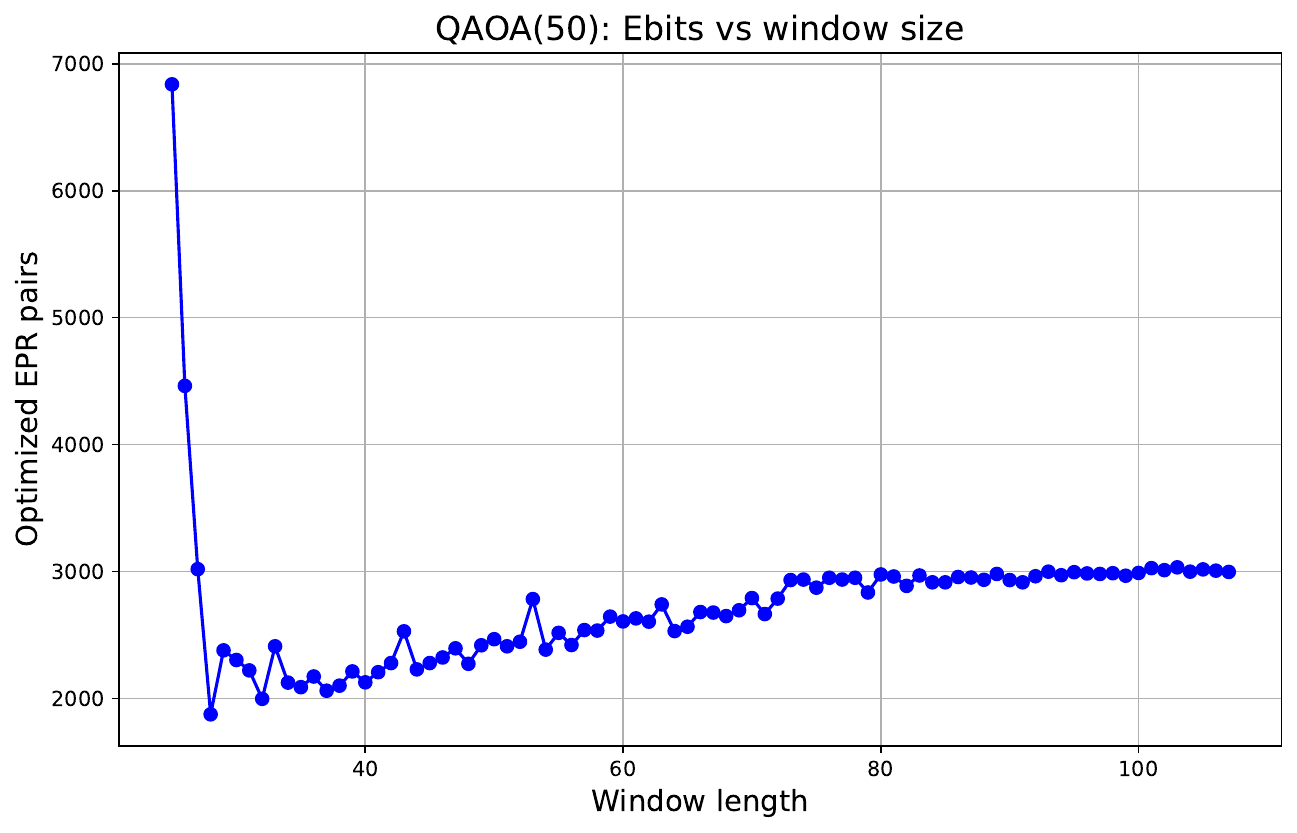} % Replace with your image file
        \caption{Impact of window size on the final EPR pairs for QAOA(50). Number of required EPR pairs vs the chosen window size for QAOA(50) when we split QAOA(50) to two processors. We use Algorithm~\ref{alg:window} and using KL algorithm as a subroutine for graph partitioning.}

    % \caption{}
    \label{fig:QAOA_plots}
\end{figure}

In this section, we discuss the numerical results obtained after implementing the Algorithm~\ref{alg:window} for various quantum circuits and compare it to the base Algorithm~\ref{algorithm:gate}. First, we consider the impact of the window size $l$ on Quantum approximate optimization algorithm (QAOA) \cite{farhi2014quantumapproximateoptimizationalgorithm} in Fig~\ref{fig:QAOA_plots}. As discussed, the algorithm is very sensitive to the chosen window size. Therefore, to optimize the EPR count, we evaluate various window sizes. For the test algorithms considered, it is generally sufficient to sweep up to one-fourth of the total number of two-qubit gates. In Fig~\ref{fig:QAOA_plots}, we plot the impact of window size on the final EPR requirement for Quantum Approximate Optimization Algorithm (QAOA) applied on 50 qubits distributed on two QPUs. We note that choosing a too small window size can be detrimental choice to make. This can be explained as the algorithm not looking far ahead into the circuit to make an optimal choice of the partition. We also see that the window size above hundred does not have much of an impact on the number of ebits. As the window size increases, the algorithm gains a more global view of the circuit’s connectivity, ultimately capturing the entire circuit structure. At a certain window size, this global perspective allows the algorithm to replicate the partitioning decisions that METIS would make, leading to convergence between the two methods.

% The plots also show the efficacy of the Algorithm~\ref{alg:window} in optimization for the number of EPR pairs compared to the base algorithm Algorithm~\ref{algorithm:gate}. Note here that for large enough window sizes, the number of ebit requirements tends to the number of ebits obtained from Algorithm~\ref{algorithm:gate}. 

\begin{figure}[ht]
    \centering
    \includegraphics[width=\linewidth]{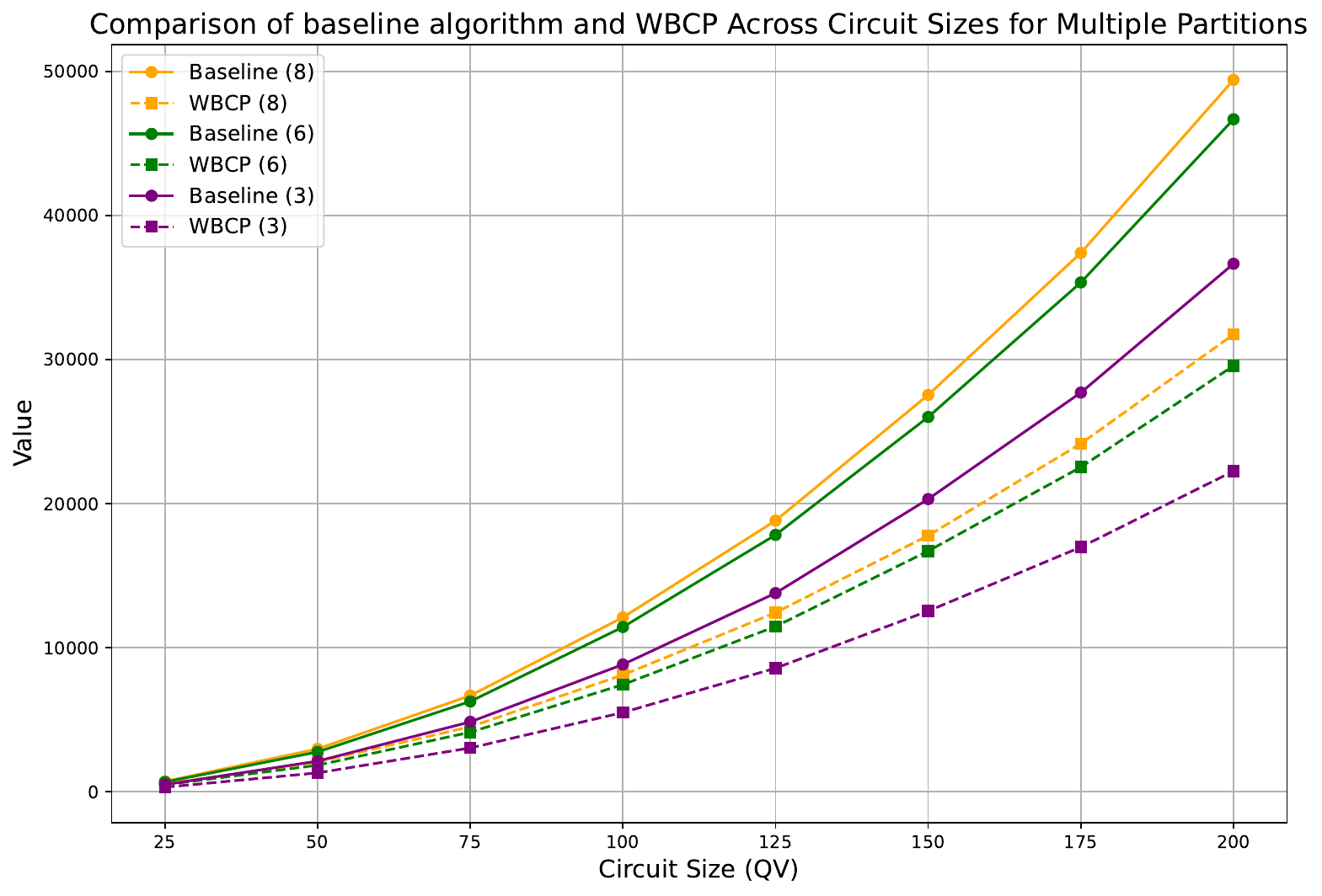}
    \caption{Comparison of the EPR pairs required to execute Quantum Volume (QV) circuits across multiple QPUs, as estimated by WBCP (Algorithm~\ref{alg:window}) and the Baseline Algorithm (Algorithm~\ref{algorithm:gate}). In the legend, the number in parentheses indicates the number of partitions. A balanced distribution of qubits across QPUs is assumed.}
    \label{fig:QV_numerics}
\end{figure}

\begin{figure}[ht]
    \centering
    \includegraphics[width=\linewidth]{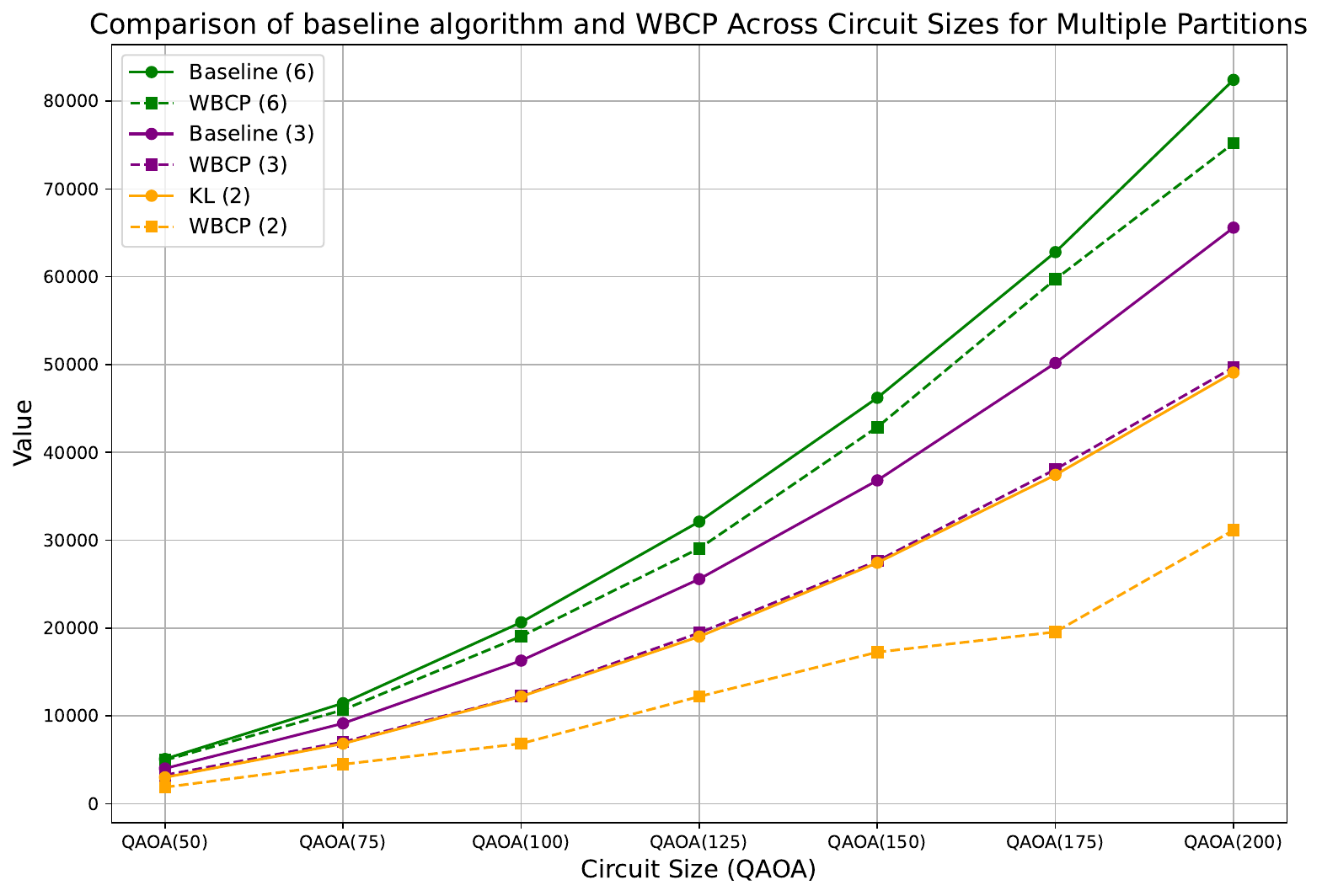}
    \caption{Comparison of the EPR pairs required to execute QAOA circuits across multiple QPUs, as estimated by WBCP (Algorithm~\ref{alg:window}) and the Baseline Algorithm (Algorithm~\ref{algorithm:gate}). In the legend, the number in parentheses indicates the number of partitions. A balanced distribution of qubits across QPUs is assumed.}
    \label{fig:QAOA_numerics}
\end{figure}

We next present the results obtained for Quantum Volume (QV) \cite{Quantum_volume} in Fig.~\ref{fig:QV_numerics}, QAOA \cite{farhi2014quantumapproximateoptimizationalgorithm} in Fig.~\ref{fig:QAOA_numerics}, and for quantum circuits described in \cite{Li23} in Table~\ref{tab:QuantumCircuits}. For Quantum Volume, we generate twenty instances of the circuit using Qiskit \cite{qiskit2024} and average the results.

A significant gap is observed between the performance of the base algorithm and the WBCP algorithm. In Fig.~\ref{fig:QV_numerics}, we analyze the Quantum Volume circuit for varying sizes. As the partition number increases for a fixed circuit size, the EPR pair requirement rises for both the base algorithm and WBCP. This behavior is expected. However, the trend of WBCP outperforming the base algorithm remains consistent. This pattern is repeated across other numerical simulations as well.

We note here that not all quantum circuits benefit from the window-based approach. As an example, for the MCMTV chain in qiskit \cite{qiskit2024}, due to it's shallow circuit depth, the base algorithm \ref{algorithm:gate}, is optimal. The usefulness of the window-based circuit depends highly on the circuit depth as well as the structure of the transpiled circuit.

\newpage

\begin{table}[!htb]
    \centering
    \caption{Comparison of EPR pairs required for circuits from \cite{Li23} split into two QPUs.}
        \centering
        
        \label{tab:QuantumCircuits}
       \begin{tabular}{|l|c|c|}
    \hline
    Circuit & KL & WBCP \\
    \hline
    Qram(20)             & 42    & 22    \\
    \hline
    Swap test(25)        & 24    & 18    \\
    \hline
    KNN(41)              & 40    & 39    \\
    \hline
    Multiplier(45)       & 339   & 198   \\
    \hline
    Square root (45)     & 5415  & 5131  \\
    \hline
    QFT(63)              & 1432  & 497   \\
    \hline
    qugan(71)            & 112   & 57    \\
    \hline
    Multiplier(75)       & 774   & 410   \\
    \hline
    QV(100)              & 6582  & 3999  \\
    \hline
    Qugan(111)           & 200   & 108   \\
    \hline
    QFT(160)             & 1560  & 192   \\
    \hline
    QFT(320)             & 1560  & 584   \\
    \hline
    Multiplier(350)      & 15644 & 4836  \\
    \hline
    Qugan(395)           & 596   & 351   \\
    \hline
    Multiplier(400)      & 21168 & 6341  \\
    \hline
\end{tabular}

    % \end{minipage}
\end{table}

\section{Distributed Quantum Fourier transform}

The Quantum Fourier Transform (QFT) \cite{QFT} is a cornerstone of many quantum algorithms, playing a crucial role in the speedup that quantum computers can achieve over classical counterparts. For example, the QFT is a key component of Shor's algorithm, which is used for factoring large integers and has significant implications for cryptography. Additionally, the QFT is employed in quantum phase estimation algorithms, which are foundational for applications in quantum chemistry, optimization, and solving linear systems of equations.

In this section, we present methods for implementing QFT in a distributed setting. While the heuristic approach in Algorithm~\ref{alg:window} is an option, we leverage the inherent structure of the QFT circuit to reduce the number of EPR pairs. By exploiting the circuit's symmetry, our method outperforms Algorithm~\ref{alg:window}.

For prior work in distributed QFT for two partitions see \cite{Yimsiriwattana_2004,Neumann2020}. \cite{Neumann2020} calculated the number of required EPR pairs to be $n/2$ for a QFT$(n)$, where $n$ refers to the number of qubits in the circuit, without considering the swapping of qubits back into original position, relegating swapping of qubits to book keeping. In this work, they used the concept of grouping multiple two-qubit gates and one EPR pair to implement this group of gates, also called as cat-entangler and cat-disentangler \cite{yimsiriwattana2004generalizedghzstatesdistributed}. 

We outline a method to implement QFT$(n)$ on two processors and extend the method to implement the QFT$(n)$ on $m \geq 3$ processors. At the end of our implementation, the positions of qubits are such that the swaps at the end of QFT circuit are not required. %We rely on teleportation of qubits instead of the method introduced in \cite{yimsiriwattana2004generalizedghzstatesdistributed}. 

\begin{figure}
    \centering
    \includegraphics[width=\linewidth]{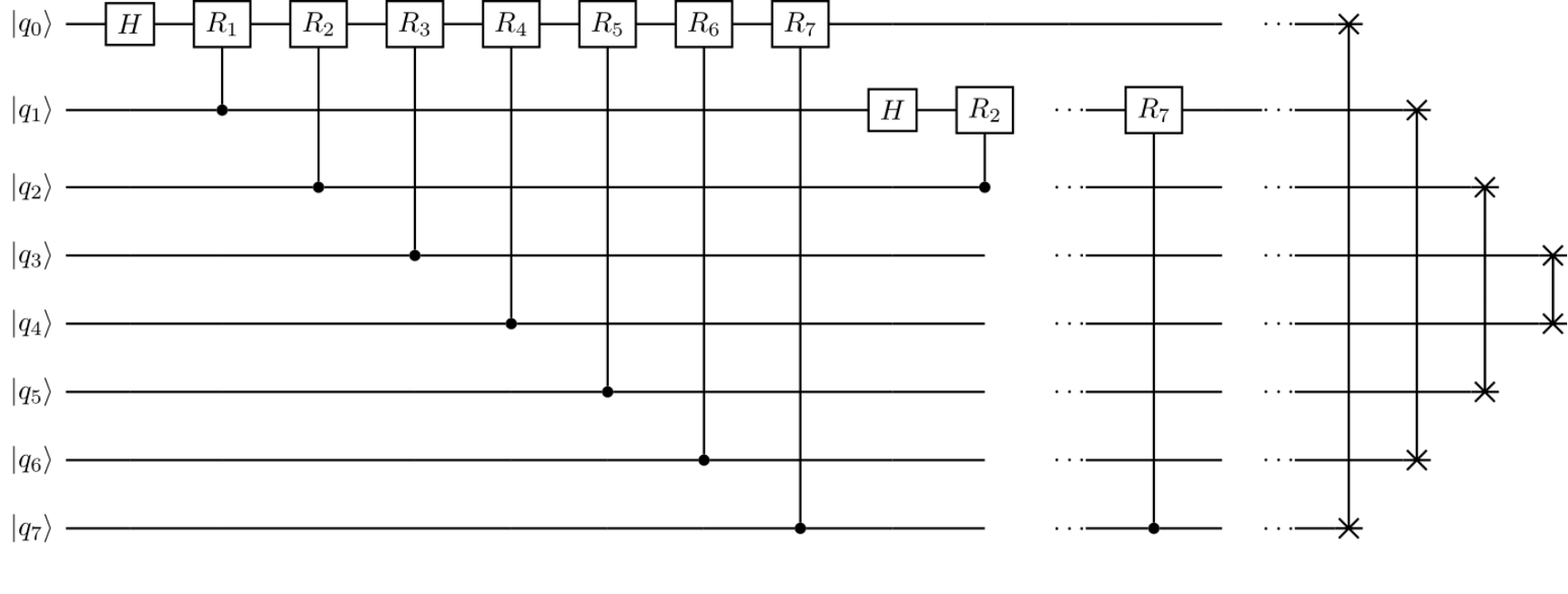}
    \caption{Implementation of QFT circuit on 8 qubits}
    \label{fig:QFT_circuit}
\end{figure}

A Quantum Fourier Transform (QFT) circuit has the following structure: For all \( i \) in \([0,n-1]\), apply a Hadamard gate \( H \) on the \( i \)-th qubit, followed by controlled rotations with \( i \) as the target qubit and \( j > i \) as the control qubit. Finally, swap qubits \( i \) and \( n-i-1\). See Fig~\ref{fig:QFT_circuit} for a QFT circuit on $8$ qubits.

We assume that the hardware natively supports controlled-rotation (CR) gates. If this is not the case, CR gates can be implemented using an ancilla-based approach or decomposed into two CX gates and a single-qubit rotation.

\subsection{QFT distributed on two QPUs}

In this section, we distribute QFT$(n)$ to two processors. We assume that both the quantum processors have $\lceil \frac{n}{2}\rceil+1$ computational qubits. The processors are also equipped with one communication qubit each. The algorithm laid out can be generalized to the case when the processors have more communication or computation qubits allowing for parallelization. The pseudocode of the method for distributing QFT on two QPUs is given in Algorithm~\ref{alg:QFT_two_processors}. In this algorithm, we label the two QPUs as QPU1 and QPU2. The qubit numbering reflects the inherent numbering in the QFT circuit and not the qubit locations in the QPU.  

Typically, the QFT reverses the order of qubits, which would require swap gates to correct the order for subsequent computations. However, in this distributed approach, as qubits are teleported between the QPUs and the corresponding gate operations required for the QFT are applied, the final qubit states emerge in the correct order without the need of additional swaps. 

The method presented in Algorithm~\ref{alg:QFT_two_processors} requires the use of $n$ EPR pairs throughout the execution of the algorithm. If the swapping order is not a concern, there exists a more efficient method described in \cite{Neumann2020} that requires only $\frac{n}{2}$ EPR pairs. Adding swapping to the protocol \cite{Neumann2020} would increase the count of EPR pairs to $3n/2$. The reason being that $n/2$ swaps would require $n$ EPR pairs.

\begin{algorithm}[H]
\caption{QFT distributed on two QPUs}
\begin{algorithmic}[1]
    \State Initialize QPU1 with qubits $[1, \frac{n}{2}]$
    \State Initialize QPU2 with qubits $[\frac{n}{2}+1, n]$
    
    \State For all $i$ in $[1,\frac{n}{2}]$, apply a $H$ gate on the $i^{\mathrm{th}}$ qubit, followed by the controlled rotations with $i$ as the target qubit and $k\geq i$ as the control qubit, where $k \in \left[i,\frac{n}{2}\right]$

    \For{$j \in [\frac{n}{2}+1, n]$}
        \State Generate entanglement between communication qubits of QPU1 and QPU2
    \State Transfer EPR pair to computational qubits
        \State Teleport qubit $j$ to computational qubit of QPU1
        \State Apply all controlled rotation gates between qubits $[1, n-j+1]$ and qubit $j$, where $j$ is the control qubit. 
        \State Generate entanglement between computational qubits of QPU1 and QPU2
        \State Teleport qubit $n-j+1$ to QPU2
        \State Apply all controlled rotation gates between qubit $n-j+1$ and qubits $[j+1, n]$, where $n-j+1$ is the target qubit. 
    \EndFor
    \State For all $i$ in $[\frac{n}{2}+1,n]$, apply a $H$ gate on the $i^{\mathrm{th}}$ qubit, followed by the controlled rotations with $i$ as the target qubit and $j\geq i$ as the control qubit.  The order is inherently reversed so there is no need of the swap gates. 

\end{algorithmic}
\label{alg:QFT_two_processors}
\end{algorithm}

\subsection{QFT distributed on multiple QPUs}

The pseudocode for distributing the Quantum Fourier Transform (QFT) across $m$ QPUs is presented in Algorithm~\ref{alg: multiple_QPUs}. The algorithm begins by dividing the $n$ qubits into $m$ blocks, with each QPU handling $k = \left\lceil \frac{n}{m} \right\rceil$ qubits. Initially, the necessary quantum gates are applied to the qubits in the first QPU. Note here that we consider $m$ even. This analysis can be extended to odd $m$. 

The algorithm then proceeds in two main phases. In the first phase, for each QPU from $i = 1$ to $m/2$, the qubits are systematically moved between QPUs, and the necessary controlled rotation (CR) gates are applied. Specifically, for each qubit $j$ in the block handled by QPU $i$, it is moved through the series of QPUs $i \rightarrow i-1 \rightarrow i-2 \ldots \rightarrow 1 \rightarrow i+1 \rightarrow i+2 \ldots  \rightarrow n-i+1$, with CR gates applied from all the qubits in QPUs $i,i-1, i-2 \ldots 1, i+1, i+2 \ldots  n-i+1$ as control and $j^{th}$ qubit as the target. During this phase, another qubit $l$ is selected from the original qubits of $n-i+1$ QPU, is moved to $i^{\textrm{th}}$ QPU.

In the second phase, for each QPU from $i = m/2$ to $1$, a similar process is followed but in reverse order. Qubits are moved back through the QPUs $i \rightarrow i-1 \rightarrow i-2 \ldots \rightarrow 1 \rightarrow i$, with CR gates applied from each qubit in the QPUs $i-1,i-2 \ldots, 1$ as the control. For this part, one can also use the cat entangler disentangler mechanism. 

This algorithm ensures that the QFT is executed efficiently across the distributed quantum system, maintaining the correct qubit order required by the QFT without the need for additional swap operations. 

% The pseudocode for distributing QFT on $m$ QPUs is given in Algorithm~\ref{alg: multiple_QPUs}. The algorithm begins by dividing the $n$ qubits into $m$ blocks, with each QPU handling $ k = \left\lfloor \frac{n}{m} \right\rfloor $ qubits.  The necessary quantum gates are applied to the first block of qubits. For each subsequent block, qubits are systematically moved between QPUs, and the necessary gates are applied as required by the QFT circuit. Specifically, qubit $j$ from block $i$ is moved through a series of QPUs, and controlled rotation (CR) gates are applied with $j$ as the control, followed by a Hadamard gate on $j$. The qubit is then moved to its final position in the corresponding reverse block $ n-i+1 $, where additional CR gates are applied. Subsequently, a different qubit $l$, selected from the original qubits in block $ n-i+1 $, is moved back through the QPUs, interacting with other qubits through CR gates and concluding with the application of an $H$ gate. This process ensures that the QFT is executed efficiently across the distributed quantum system. A key feature of this algorithm is its ability to maintain the correct qubit order required by the QFT without the need for additional swap operations. 
\begin{algorithm}[H]
\caption{Distributed QFT for Multiple QPUs}
\begin{algorithmic}[1]
    \State \textbf{Input:} Number of qubits $n$, number of QPUs $m$
    \State Divide $n$ qubits into $m$ QPU, each with $k = \left\lceil \frac{n}{m} \right\rceil$ qubits.
    \For{$i = 1$ to $m/2$}
    \State  Apply the gates in the $i$ QPU.
        \For{$j = 0$ to $k-1$}
            
            \State Move qubit $j$ of $i$ QPU  from $i \rightarrow i-1 \rightarrow i-2 \rightarrow 1 \rightarrow  i+1 \rightarrow i+2 \ldots  \rightarrow m-i+1$
            
            \State Apply CR gates with qubit $j$ as the target in QPUs   $ i-1, i-2, 1, i+1 , i+2 \ldots  , m-i+1$.

            \State Select qubit $l$ from the original qubits in block $m-i+1$ 
            \State Move qubit $l$ to QPU $i$. Apply CR gate with $l$ as control in QPU $i$. 
        
        \EndFor
    \EndFor
    \For{$i = m/2$ to $1$}
    \State Apply all gates in $i$ QPU.
        \For{$j = 0$ to $k-1$}
            \State Move qubit $j$ from QPU $i \rightarrow i-1\rightarrow i-2  \ldots \rightarrow 1 \rightarrow i$
            
            \State Apply CR gates with qubit $j$ as the target in QPUs   $ i-1, i-2, i-3 \ldots , 1$.

        \EndFor
    \EndFor
\end{algorithmic}
\label{alg: multiple_QPUs}
\end{algorithm}

% \begin{algorithm}
% \caption{Distributed QFT for Multiple QPUs}
% \begin{algorithmic}[1]
%     \State \textbf{Input:} Number of qubits $n$, number of QPUs $m$
%     \State Divide $n$ qubits into $m$ QPU, each with $k = \left\lceil \frac{n}{m} \right\rceil$ qubits.
%     \State Apply the gates in the first QPU.
    
%     \For{$i = m/2$ to $1$}
%         \For{$j = 0$ to $k-1$}
%             \State Move qubit $j$ from QPU $i \rightarrow i-1 \rightarrow \ldots \rightarrow 1 \rightarrow n-i+1$
            
%             \State Apply CR gates with qubit $j$ as the control in QPUs $i \rightarrow i-1 \rightarrow \ldots \rightarrow 1$
%             \State Apply $H$ to the $j$ qubit. 
%             \State In the $n-i+1$ block, apply CR gates from the original qubits of the block to qubit $j$
            
%             \State Select qubit $l$ from the original qubits in block $n-i+1$ (where $l \neq j$)
%             \State Move qubit $l$ from QPU $n-i+1 \rightarrow n-i \rightarrow \ldots \rightarrow i+1 \rightarrow i-1 \rightarrow \ldots \rightarrow 1 \rightarrow i$
%             \State Apply CR gates with qubit $l$ as the control to the remaining qubits in each block
%             \State Apply $H$ to the $l$ qubit and CR gates from the new qubits in the block $i$.
%         \EndFor
%     \EndFor
% \end{algorithmic}
% \label{alg: multiple_QPUs}
% \end{algorithm}

\subsubsection{EPR calculation}

The total number of EPR pairs required by the Algorithm~\ref{alg: multiple_QPUs} is determined by considering the movement and interaction of qubits across QPUs.

For the first for loop, we can calculate the EPR pairs as 
% \begin{align}
%     \text{Total EPR pairs first loop} &\leq k\sum_{i=1}^{m/2} (m+2-2i +i -1) \\
%     &= k\sum_{i=1}^{m/2} (m+1-i)
% \end{align}
\begin{align}
    \text{Total EPR pairs first loop} &\leq 
    k\sum_{i=1}^{m/2} (m+1-i)
\end{align}
To understand this, observe that the number of blocks visited by any qubit from the \(i^{\textrm{th}}\) block is \(m-i\). Then, qubits from \(m-i\) blocks must return to the original \(i^{\textrm{th}}\) block.

If we use CAT entangler and disentangler for the second loop, we can calculate the EPR pairs as 
\begin{equation}
    \text{Total EPR pairs second loop} \leq k\sum_{i=1}^{m/2} (i-1)  
\end{equation}

Then, the total EPR pairs are
\begin{equation}
    \text{Total EPR pairs} \leq \frac{km^2}{2}
\end{equation}

The inequality in the above equation comes from the fact that the number of qubits in each QPU are upper bounded by $k$.

% For $i \in [2, \frac{m}{2}]$, the number of EPR pairs needed can be calculated as:

% \begin{equation}
% \text{Total EPR pairs} = k \times \left(\frac{n(n+1)}{2} - 1\right)
% \end{equation}

% where $k = \left\lceil \frac{n}{m} \right\rceil$ is the number of qubits in each block. To see this, note that any qubit in $i^{\textrm{th}}$ block has to be teleported $i$ times. Each block has an upper bound of $k$ qubits. 

We also generalize the distributed QFT algorithm presented in \cite{Neumann2020} to accommodate multiple QPUs. Let us consider $m$ QPUs or blocks. Each block $i \in [1,m]$ must interact with $m-i$ other blocks to apply the remote controlled rotation (CR) gates, where each block contains $k$ qubits. Every interaction results in the exchange of $k$ EPR pairs. Consequently, the total number of EPR pairs required can be calculated as $\frac{m(m-1)k}{2}$. While this approach does offer better EPR consumption compared to the one in Algorithm~\ref{alg: multiple_QPUs}, it does not account for the final swap gates. For a fair comparison, we account for the swap gates in the extension of Algorithm in \cite{Neumann2020}. The EPR pair consumed by including the swap gate is $\frac{m(m-1)k}{2}+n = \frac{km^2}{2} + \frac{km}{2}$, where we assume $n = km$. We then find that the algorithm introduced in Algorithm~\ref{alg: multiple_QPUs} uses less number of EPR pairs in comparison. 

We summarize the results as follows: given an \( n \)-qubit QFT circuit partitioned into \( m \) QPUs, the number of EPR pairs required by the described method is \( \frac{nm}{2} \). In contrast, the number of EPR pairs required when generalizing the method from \cite{Neumann2020} is \( \frac{nm}{2} + \frac{n}{2} \).

\section{Network topology considerations}

We envision a network of QPUs connected by fibers and switches. While it is possible to establish entanglement between any pair of QPUs, the efficiency of these connections can vary due to factors like the number of intermediary switches or the potential loss in the fiber. This means that, in practice, some QPU connections may be more resource-intensive than others, despite the theoretical possibility of all-to-all connectivity. An example of this kind of architecture can be found in the proposed quantum data center in \cite{DQC_cisco}.

In our model, the actual physical QPUs present in a data center are represented as a graph $G \equiv (V, E)$, where each node corresponds to a physical QPU, and each edge represents a connection between these QPUs. The weight assigned to each edge reflects the cost associated with that specific connection, considering factors such as distance, the number of switches, and potential signal loss.

When analyzing a quantum circuit, we can use Algorithm~\ref{alg:window} or any other circuit partitioning methods, to determine the number of EPR pairs required between the QPUs. These requirements can be mapped onto a separate graph $G' \equiv (V',E')$, where the nodes represent the QPUs involved in the circuit execution, and the edge weights indicate the number of EPR pairs needed. This graph $G'$ serves as an abstract representation of the connectivity demands based on the quantum circuit, distinct from the physical network graph $G$.

We map $G'$ onto $G$, determining the optimal placement of QPUs within the network. The cost function is calculated as the sum of the weights associated with the links between the physical QPUs—determined by the number of switching events and loss in the fiber—multiplied by the communication requirements (e.g., the number of EPR pairs) between them. Intuitively, the cost function encourages QPUs that require a higher number of EPR pairs to be placed on physical QPUs connected by low-loss links, thereby reducing the overall demand on the network. We formulate the above optimization as an integer linear program in next section.

\subsection{ILP formulation}\label{sec:ILP_formulation}
In this section, for a graph $G$, we use the notation $E(G)$ to denote the edge set of the graph $G$ and $V(G)$ to denote the vertex set. Given an edge $i \in E(G)$, we denote the pair of vertices connected by $i$ as $V(i) = (u,v)$, where $u, v \in V(G)$.

Let $w_i$ be the weight of the $i^{\mathrm{th}}$ edge in $E(G)$, and $l_j$ be the weight of the $j^{\mathrm{th}}$ edge in $E(G')$.
We can consider the weight between the two nodes in $G$ as the \textit{infidility} of the link connecting the QPUs. Thus, higher weight corresponds to noisier link. The $l_j$ can be considered as the number of EPR pairs required from the edge $j$. Let $x_{ij}$ be a binary variable that equals one if the $i$ edge in $E$ corresponds to $j$ edge in $E'$. Let $y_{uv}$ be the binary variable that equals one if the $u$ node in $V$ corresponds to the $v$ node in $V'$. 

Then, we can formulate the question of minimizing the usage of the noisier link in $G'$ as the following quadratic optimization problem: 
\begin{align}
 &\min_{x_{ij}} \sum_{ij} x_{ij} w_i l_j\\
 \quad s.t. &\sum_{v \in V(G')} y_{uv} = 1 \quad \forall u \in  V(G)\\
 \quad &\sum_{u \in V(G)} y_{uv} = 1 \quad \forall v \in V(G')\\
 \quad &\sum_{j \in E(G')} x_{ij} = 1 \quad \forall i \in E(G)\\
  \quad &\sum_{i \in E(G)} x_{ij} = 1 \quad \forall j \in E(G')\\
\quad x_{ij} &= y_{u_1v_1} \cdot y_{u_2v_2} + y_{u_1v_2} \cdot y_{u_2v_1} \quad\nonumber \\&\forall i \in E(G), \forall j \in E(G'), \text{ where } u_1, u_2 \in V(i), \nonumber \\ &\text{ and } v_1, v_2 \in V(j)
\label{eq: quadratic_constraint}
\end{align}

The first equation reflects the cost function that we seek to minimize. The second equation constraints the search space such that each vertex in $V(G')$ is mapped to exactly one vertex in $V(G)$. The third equation implies that each vertex in $V(G)$ is mapped to exactly one vertex in $V(G')$. The fourth equation implies that each edge in $E(G')$ is mapped to exactly one edge in $E(G)$. The fifth equation implies that each edge in $E(G)$ is mapped to exactly one edge in $E(G')$. The last quadratic equation ensures that the mapping of edges between two graphs is consistent with the node assignments, meaning that an edge between two nodes in one graph is correctly mapped to an edge between the corresponding nodes in the other graph.

Next, we map the quadratic constraint in Equation~\ref{eq: quadratic_constraint} to a linear constraint by introducing auxillary variables $z1_{ij}, z2_{ij}$. 

% The original quadratic constraint is given by:
% \begin{align}
% x_{ij} = y_{u_1 v_1} \cdot y_{u_2 v_2} + y_{u_1 v_2} \cdot y_{u_2 v_1}, \quad \forall i \in E(G'), \forall j \in E(G)
% \end{align}

To linearize this constraint, we introduce auxiliary binary variables \( z1_{ij} \) and \( z2_{ij} \) such that:
\begin{align}
z1_{ij} &= y_{u_1 v_1} \cdot y_{u_2 v_2} \quad \forall i \in E(G), \forall j \in E(G'), \nonumber\\&\text{ where } u_1, u_2 \in V(i), \text{ and } v_1, v_2 \in V(j)\\
z2_{ij} &= y_{u_1 v_2} \cdot y_{u_2 v_1}\quad \forall i \in E(G), \forall j \in E(G'), \nonumber\\&\text{ where } u_1, u_2 \in V(i), \text{ and } v_1, v_2 \in V(j)
\end{align}

The quadratic constraint can be enforced with the following linear constraints:
\begin{align}
z1_{ij} &\leq y_{u_1 v_1}, \quad \forall i \in E(G), \forall j \in E(G'), \nonumber\\&\text{ where } u_1 \in V(i), \text{ and } v_1 \in V(j) \\
z1_{ij} &\leq y_{u_2 v_2},\quad \forall i \in E(G), \forall j \in E(G'), \nonumber\\&\text{ where } u_2 \in V(i), \text{ and } v_2 \in V(j) \\
z1_{ij} &\geq y_{u_1 v_1} + y_{u_2 v_2} - 1, \quad \forall i \in E(G), \forall j \in E(G'), \nonumber\\&\text{ where } u_1, u_2 \in V(i), \text{ and } v_1, v_2 \in V(j)
\end{align}
\begin{align}
z2_{ij} &\leq y_{u_1 v_2}, \quad \forall i \in E(G), \forall j \in E(G'), \nonumber\\&\text{ where } u_1 \in V(i), \text{ and } v_2 \in V(j) \\
z2_{ij} &\leq y_{u_2 v_1}, \quad \forall i \in E(G), \forall j \in E(G'), \text{ where } u_2 \in V(i), \nonumber\\&\text{ and } v_1 \in V(j) \\
z2_{ij} &\geq y_{u_1 v_2} + y_{u_2 v_1} - 1, \quad \forall i \in E(G), \forall j \in E(G'), \nonumber\\&\text{ where } u_1, u_2 \in V(i), \text{ and } v_1, v_2 \in V(j)
\end{align}

Finally, we replace the original constraint with the following linear constraint:
\begin{align}
x_{ij} = z1_{ij} + z2_{ij}, \quad \forall i \in E(G), \forall j \in E(G')
\end{align}

The formulation does result in an increased number of variables in the optimization problem. However, the linearity of both the objective function and the constraints enables us to leverage ILP optimization tools effectively. 

It is possible to incorporate symmetry in the above formulation to reduce the number of variables. We leave that for future explorations.

%For details, see Appendix~\ref{appendix_ILP_symmetry}.

% In the last constraint, $i$ is an edge between $u_1, u_2$ and $j$ is an edge between $v_1,v_2$.

% Reformulate the last constraint as 
% \begin{equation}
%     y_{u_1v_1} + y_{u_2v_2} +  y_{u_2v_1} + y_{u_1v_2} = x_{ij}+1
% \end{equation}

% Then, we can formulate the question of minimizing the usage of the noisier link in $G'$ as the following minimization problem: 
% \begin{align}
%  &\min_{x_{ij}} \sum_{ij} x_{ij} w_i l_j\\
%  \quad s.t. &\sum_{v \in V(G)} y_{uv} = 1 \quad \forall \textrm{u in } V(G')\\
%  \quad &\sum_{j \in E(G)} x_{ij} = 1 \quad \forall \textrm{i in } E(G')\\
%  \quad & y_{u_1v_1} + y_{u_2v_2} +  y_{u_2v_1} + y_{u_1v_2} = x_{ij}+1\forall \textrm{ i} \in \ E_G \textrm{ and } j \in E_{G'}
% \end{align}
% In the last constraint, $i$ is an edge between $u_1, u_2$ and $j$ is an edge between $v_1,v_2$.

% \bibliographystyle{IEEEtran}
% \bibliography{main}

\subsection{Numerical Results}
In this section, we compare the solution proposed in Section~\ref{sec:ILP_formulation} to the approach of randomly assigning QPUs in a quantum data center. 

We first start with generating the weights $w_i$ for the graph $G$, representing the infidelity of the EPR pair transported along the edge $i$ connecting the two nodes. To generate these weights, an upper triangular matrix was generated using a beta distribution. The two parameters of the distribution were set equal and then varied to control the skewness of the distribution. The parameter is henceforth denoted by $\beta >0$. 

Next, we generate the graph $G'$, which corresponds to the demand of the EPR pairs between the QPUs. This graph is created by assigning weights to the edges that are biased towards extreme values, reflecting the varying communication demands between different pairs of QPUs.

\begin{figure}[h]
    \centering
    \includegraphics[width=\linewidth]{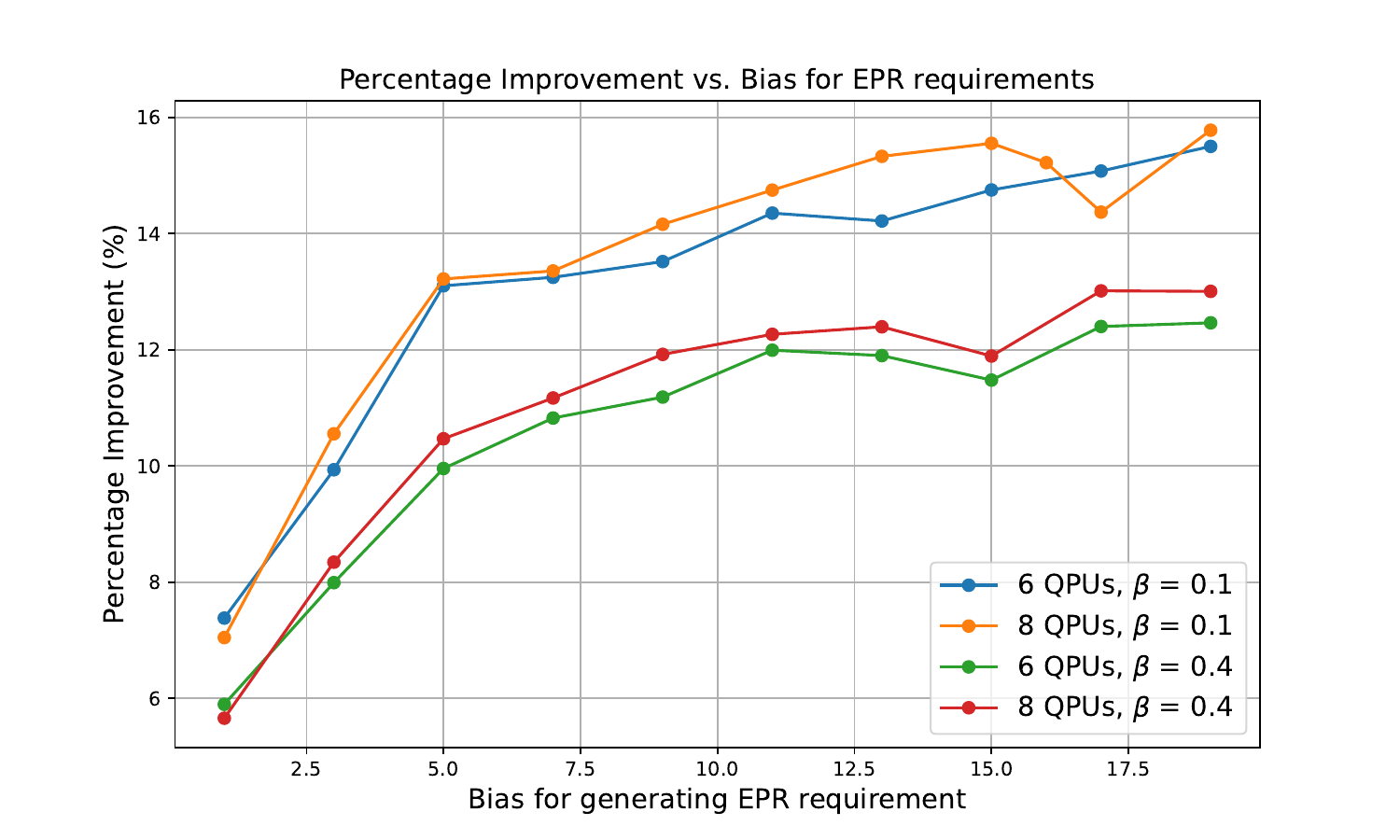}
    \caption{In this figure, we plot the percentage improvement of the ILP objective function against the bias of the probability distribution generating the EPR demands (weights for \( l_i \)), shown on the x-axis, and the bias of the probability distribution ($\beta$) generating the infidelity of the connecting links (weights for \( w_i \)).}
    \label{fig:bias_percentage_improv}
\end{figure}

We note here that were the weights $w_i$ and $l_i$ generated randomly, the optimization does not yield any advantage. We vary the parameter controlling the skewness of the distributions generating the weights of the graph $G$ and $G'$, and plot the results in Figure~\ref{fig:bias_percentage_improv}. We see as the skewness of the two distribution increases, the advantage that we gain from the optimization also increases. 

\begin{figure}[ht]
    \centering
    \begin{subfigure}{0.45\textwidth}
        \centering
        \includegraphics[width=\textwidth]{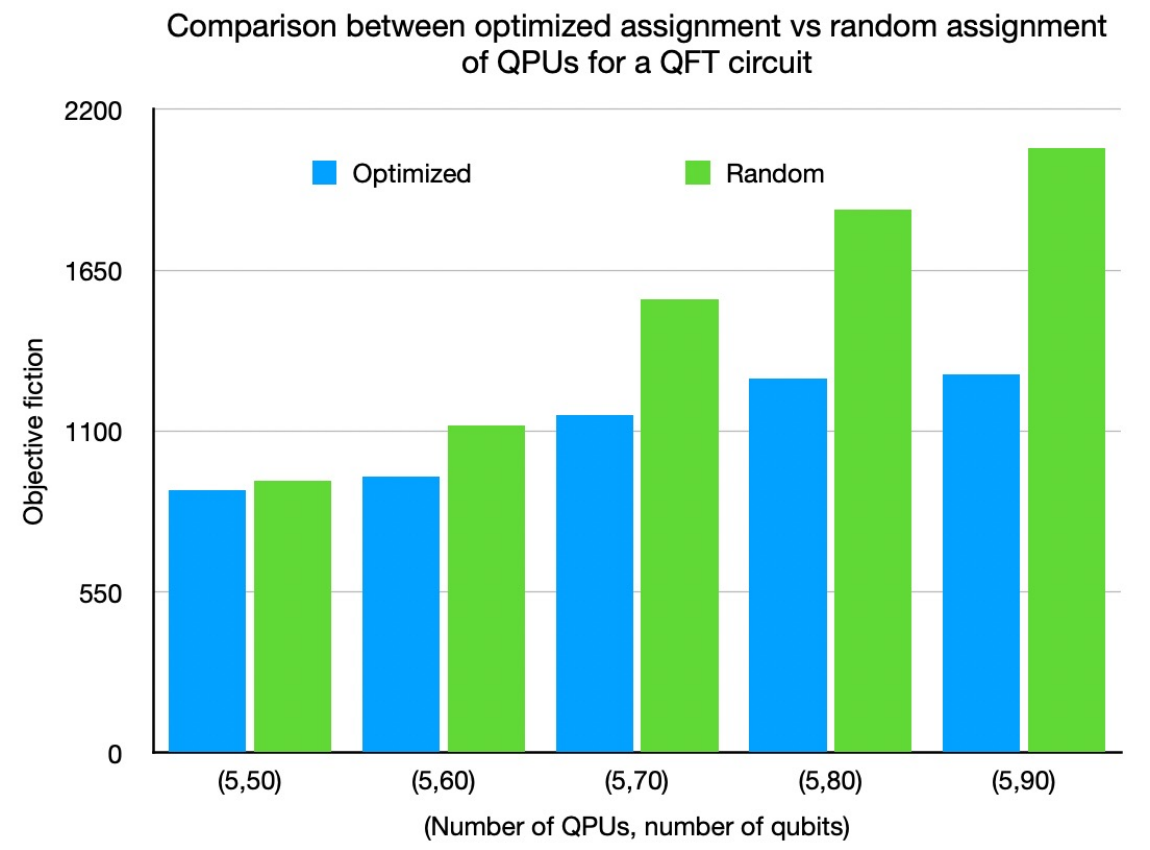}
    \end{subfigure}
    \caption{In this figure, we compare the value of the objective function for a random QPU assignment with that of an optimized QPU assignment for QFT circuits.}
    \label{fig:qft_assignment}
\end{figure}

Additionally, Figure~\ref{fig:qft_assignment} showcases the application of the optimization approach to the Quantum Fourier Transform (QFT) circuit. The QFT is analyzed for various qubit counts and distributed equitably across QPUs. The Qiskit compilation of QFT generates an optimized circuit, which in turn impacts the demands for entanglement between QPUs. Due to circuit optimization, the EPR pair demands become skewed across QPUs rather than being uniformly distributed.
 By applying the base algorithm, the EPR pairs between each pair of QPUs are obtained, and the topology graph is generated using a skewed \(\beta\) distribution. 
This imbalance, combined with the inherent skewness in the entanglement demands, leads to the optimization giving an improvement over the random assignment.

% \subsection{Numerical Results} Figure~\ref{fig:optimization_QPUs} demonstrates the application of the ILP method for QPU assignment, aimed at minimizing the cost function \(\sum_{i} w_i l_i\). Here, \(w_i\) represents the weight of the \(i^{\textrm{th}}\) edge in the topology graph \(G\) (defined by the network parameters), and \(l_i\) represents the weight of the \(i^{\textrm{th}}\) edge in the demand graph \(G'\). The results indicate that the optimized assignment consistently outperforms random assignment. For the numerical analysis, \(w_i\) and \(l_i\) are generated using a skewed \(\beta\) distribution, and the cost function is subsequently solved.

\section{Conclusion}

In conclusion, this work proposes a novel approach to optimizing qubit allocation in quantum data centers. By utilizing graph partitioning techniques and integer linear programming, the algorithm efficiently minimizes the EPR pairs required to execute quantum circuits across multiple QPUs. This method not only tackles the challenge of reducing overall EPR pair usage but also strategically alleviates the burden on low-fidelity and low-capacity links. Additionally, we present a technique to optimize EPR demands for a QFT circuit distributed across multiple QPUs. Future work will expand this model to accommodate more complex quantum data centers and refine the implementation of two-qubit gates through gate packing \cite{Wu2023}.

% \acknowledgements
% 
\acknowledgements
\vspace{-10pt} % Adjust as needed
The authors acknowledge insightful discussions with Stephen DiAdamo and Troy Sewell.

\bibliography{main.bib}

% \printbibliography

\end{document}